\documentclass[%
reprint,
superscriptaddress,
 amsmath,amssymb,
aps,
]{revtex4-1}

\usepackage{graphicx}
\usepackage{dcolumn}
\usepackage[dvipsnames]{xcolor}
\usepackage{bm}
\usepackage[export]{adjustbox}

\begin{document}


\title{Homophily-based social group formation in a spin-glass self-assembly framework}

\author{Jan Korbel}
 \affiliation{Section for the Science of Complex Systems, CeMSIIS, Medical University of Vienna, Spitalgasse 23, A-1090, Vienna,
Austria}
\affiliation{Complexity Science Hub Vienna, Josefst\"{a}dterstrasse 39, A-1080, Vienna, Austria}

\author{Simon D. Lindner}
 \affiliation{Section for the Science of Complex Systems, CeMSIIS, Medical University of Vienna, Spitalgasse 23, A-1090, Vienna,
Austria}
\affiliation{Complexity Science Hub Vienna, Josefst\"{a}dterstrasse 39, A-1080, Vienna, Austria}

\author{Tuan Minh Pham}
 \affiliation{Section for the Science of Complex Systems, CeMSIIS, Medical University of Vienna, Spitalgasse 23, A-1090, Vienna,
Austria}
\affiliation{Complexity Science Hub Vienna, Josefst\"{a}dterstrasse 39, A-1080, Vienna, Austria}
\affiliation{Niels Bohr Institute, Blegdamsvej 17, 2100 Copenhagen, Denmark}

\author{Rudolf Hanel}
 \affiliation{Section for the Science of Complex Systems, CeMSIIS, Medical University of Vienna, Spitalgasse 23, A-1090, Vienna,
Austria}
\affiliation{Complexity Science Hub Vienna, Josefst\"{a}dterstrasse 39, A-1080, Vienna, Austria}

\author{Stefan Thurner}
\email{stefan.thurner@meduniwien.ac.at}
\affiliation{Section for the Science of Complex Systems, CeMSIIS, Medical University of Vienna, Spitalgasse 23, A-1090, Vienna,
Austria}
\affiliation{Complexity Science Hub Vienna, Josefst\"{a}dterstrasse 39, A-1080, Vienna, Austria}
\affiliation{Santa Fe Institute, 1399
Hyde Park Road, Santa Fe, NM, 87501, USA}
\date{\today}

\begin{abstract}
Homophily, the tendency of humans to attract each other when sharing similar features, traits, or opinions has been identified as one of the main driving forces behind the formation of structured societies. Here we ask to what extent homophily can explain the formation of social groups, particularly their size distribution. We propose a spin-glass-inspired framework of self-assembly, where opinions are represented as multidimensional spins that dynamically self-assemble into groups; individuals within a group tend to share similar opinions (intra-group homophily), and opinions between individuals belonging to different groups tend to be different  (inter-group heterophily). We compute the associated non-trivial phase diagram by solving a self-consistency equation for 'magnetization' (combined average opinion). Below a critical temperature, there exist two stable phases: one ordered with non-zero magnetization and large clusters, the other disordered with zero magnetization and no clusters. The system exhibits a first-order transition to the disordered phase. We analytically derive the group-size distribution that successfully matches empirical group-size distributions from online communities.
\end{abstract}

\maketitle


Structure-forming systems form an important class of complex systems \cite{thurner2018}. They are ubiquitous in natural and social systems, ranging from atoms forming molecules, polymers, colloids and micelles to people forming structured societies. The theory of  \emph{self-assembly} \cite{whitesides2002} describes the emergence of higher-order structures from elementary components. Applications include molecular self-assembly \cite{whitesides1991}, lipid bilayers and vesicles \cite{israelachvili1977}, microtubules and molecular motors \cite{aranson2006theory}, Janus particles \cite{fantoni2011,walther2013}, other types of patchy particles \cite{kern2003}, and RNA self-assembly \cite{grabow2014}. The thermodynamics of self-assembled systems can be described sufficiently well with the grand-canonical ensemble for large systems. This is no longer true for small systems consisting of dozens or hundreds of particles. Correct results are obtained from the canonical ensemble with an appropriate correction to the entropic functional that correctly accounts for the statistics of structure formation \cite{korbel2021}.

Social group structures emerge from interactions between individuals. While traditional approaches explore social group formation under endogenous factors \cite{hechter1978,gueron1995,zeggelink1996}, more recent works attempt to explain its structures as a consequence of opinion formation \cite{Klimek2016,Schweitzer2019,Schweitzer2020,schweitzer2021}. Within this framework, groups are considered as clusters of homogeneous agents whose opinions evolve under the joint effects of \emph{structural balance} -- the tendency to resolve tension in unbalanced triadic interactions \cite{heider1946} and  \emph{homophily} -- the preference of like-minded individuals to cluster \cite{mcpherson2001,huber2017}. Both approaches can explain the fragmentation of society into well-connected groups of uniform opinions, sometimes referred to as  \emph{echo chambers} \cite{marvel2009,CURRARINI201618,belaza2017statistical,gorski2020,minh2020,minh2021}.   Spin glass Hamiltonians have been used on static social interaction networks to quantify the amount of social stress of the entire society \cite{minh2021}, or that of each individual \cite{Tuan22}. Social stress plays the role of energy and measures opinion-similarity between individuals.
Spin glass models were extensively studied on various network topologies, including fully-connected  \cite{griffiths1966relaxation,GALAM,kochmanski2013curie,colonna2014anomalous}, Barab\'{a}si-Albert  \cite{bianconi2002,merger2021global}, small-world \cite{pekalski2001}, more general \cite{dorogovtsev2002}, and co-evolutionary, dynamic networks \cite{BielyHanelThurner,sornette2014,raducha2018}; see \cite{dorogovtsev2008} for a review. { A similar idea of considering group formation as a way to maximize payoff through local homophilic interaction has led the author in \cite{javarone2017evolutionary}  to the observation of a transition between the "group"- and the "individual" phases upon varying the ratio between individual payoff and group payoff.
}

Obviously, the formation of friendship groups from individuals that randomly encounter each other cannot be realistically described on static networks. The theory of self-assembly offers an attractive alternative that could explain the endogenous emergence of
social groups.
To capture the interplay of opinion dynamics and group formation, the assumption of a stochastic rule for establishing social ties based on the similarity of opinions is reasonable.

To realize such a model, we assume an attractive interaction between individuals based on the proximity of their opinions \cite{Tuan22}. Opinions are represented by Ising-like spin vectors in $G$ dimensions, each dimension corresponding to one binary opinion on a specific topic; the more aligned these vectors, the stronger the attractive force and the more likely they will form a friendly social tie. The main idea behind the model is that people tend to form friendship groups with like-minded individuals. They can also form hostile relations with individuals -- typically from other groups. Entertaining a friendship relation with an individual with a drastically different opinion creates \emph{social stress}. To reduce it, one can either change opinions or move to another group.

To overcome the main limitation of previous models --- the pre-defined social network topology ---  we assume that individuals create the social network by dynamically interacting with each other and forming social links stochastically. We assume that every individual has a typical (average) number of positive connections within their group. At times, with a certain probability, people meet individuals from other groups. Links between individuals that belong to different groups are typically negative since they tend to have non-aligned opinion vectors.
We assume that the probability of establishing a new (positive or negative) link between two individuals is proportional to the number of links both individuals have. The new link is positive if two individuals share more than half of their opinions; the link is negative if the majority of opinions are different.
In the model, the social network emerges dynamically; only the local quantities such as each individual's typical degree (the number of social interactions) are needed as an input. The resulting equilibrium group-size distribution can then be derived using the theory of self-assembly \cite{korbel2021}. The distribution depends on the ``temperature'', $T$, represents the willingness to change one's opinion or to change the group.
We study the phase structure (location of tipping points) of the model and compute the group-size distribution that is compared to real data.
We confirm all analytical findings with straightforward Monte Carlo (MC) simulations. The core of the model is a social stress function (Hamiltonian) that every individual tries to minimize by either changing their opinion or their group membership.

\emph{Self-assembly of spin glass.---} Let us consider $n$ individuals with $g$ binary opinions (spin vectors).
We denote the $j$-th opinion of individual $i$ by $s_i^j \in \{-1,1\}$. The spin vector of the $i$-th individual is $\mathbf{s}_i = \{s_i^1,\dots,s_i^g\}$. We define the homophily between two individuals
as the normalized dot-product, $\mathbf{s}_i \cdot \mathbf{s}_j = \frac{1}{g} \sum_{l=1}^g s_i^l s_j^l$. Individuals can form clusters of any size, $k \in \{1,\dots,n\}$. We denote the number of groups of size $k$ by $n^{(k)}$; these fulfil $\sum_{k=1}^n k n^{(k)} = n$, 
where $\boldsymbol{\cdot}^{(k)}$ denotes the dependence of a quantity on a given group size, $k$.
A group of size $k$ is given by $\mathcal{G}^{(k)}=\{i_1,\dots,i_k\}$. Following \cite{Tuan22}, we define the {\em group Hamiltonian} as
\begin{eqnarray}
H(\mathbf{s}_{i_1},\dots,\mathbf{s}_{i_k}) &:=& - \phi \, \frac{J}{2} \! \sum_{ij \in \mathcal{G}^{(k)}}  A_{ij} \, \mathbf{s}_i \cdot \mathbf{s}_j \nonumber\\
&&+ (1-\phi) \, \frac{J}{2} \! \sum_{i \in \mathcal{G}^{(k)}, j \notin \mathcal{G}^{(k)}}  A_{ij} \, \mathbf{s}_i \cdot \mathbf{s}_j
\nonumber\\
&&- h^{(k)} \sum_{i \in \mathcal{G}^{(k)}} \mathbf{s}_i \cdot \mathbf{w}
\end{eqnarray}
where $J>0$ is the coupling constant, $A_{ij}$ is the (dynamical) adjacency matrix of the underlying interaction network. The first term corresponds to homophilic intra-group interactions.
The second term captures the inter-group interactions.
The parameter, $\phi$, weights the relative importance of intra-group and inter-group stress.
The last term corresponds to an external bias caused, e.g., by the mass media, $h^{(k)}$ is the local external field that encodes the strength of that bias, and $\mathbf{w}$ is a weight vector, measuring sensitivity to that bias. We take $\mathbf{w} = \{1,\dots,1\}$.
{ As discussed in \cite{Tuan22}, for $G=1$ the model reduces to the spin model of Mattis type \cite{mattis1976} (also used in \cite{Facchetti2011,Singh2014}). This model has no frustration on effective spins $\mathbf{\tau}_i = \mathbf{\epsilon}_i \mathbf{s}_i$, where $\epsilon_{i} \epsilon_{j} =1$ for $j\in \mathcal{G}$ and $\epsilon_{i} \epsilon_{j} =-1$ if $j\notin \mathcal{G}$. Thus, the effective Hamiltonian in terms of $\tau_i$ is the usual Ising Hamiltonian with no negative interactions. This is, however, not possible for $G>1$, and therefore we inevitably end with frustrated interactions (at least for some opinions).}

The relative number of clusters of size $k$ is $\wp^{(k)} = n^{(k)}/n$. The equilibrium group-size distribution can be expressed as \cite{korbel2021}
\begin{equation}
\wp^{(k)} = \Lambda^k \mathcal{Z}^{(k)}
\end{equation}
where $\mathcal{Z}^{(k)} = \frac{n^{k-1}}{k!} \sum_{\mathbf{s}_{i_1},\dots,\mathbf{s}_{i_k}} e^{-\beta H(\mathbf{s}_{i_1},\dots,\mathbf{s}_{i_k})}$ is the partition function of a group with size $k$, $\beta=\frac{1}{k_B T}$ is the inverse temperature (using $k_B=1$), and $\Lambda$ is the normalization obtained from \mbox{$\sum_{k=1}^n k \wp^{(k)} = \sum_{k=1}^n k \,\Lambda^k \mathcal{Z}^{(k)} = 1$}, which is a polynomial equation in $\Lambda$ of order $n$. The number of groups per individual is $M = \sum_{k=1}^n n^{(k)}/n = \sum_{k=1}^n \wp^{(k)}$ and the average group size is $C = \sum_{k=1}^n  k \, n^{(k)}/\sum_{k=1}^n n^{(k)} = 1/M$. Choosing $\mathbf{w}=(1,\dots,1)$,  the average opinion vector of group $\mathcal{G}^k$ is defined as $\mathbf{m}^{(k)} =  \sum_{i \in \mathcal{G}^{(k)}}  \langle \mathbf{s}_i \rangle$. The average weighted opinion $m^{(k)} = \mathbf{m}^{(k)} \cdot \mathbf{w}$ can be expressed as $m^{(k)} = - \frac{1}{\beta} \frac{\partial \log \mathcal{Z}^{(k)}}{\partial h^{(k)}}$;
the total magnetization divided by the number of individuals is therefore $m = \sum_k \wp^{(k)} m^{(k)}$.

\emph{Simulated annealing.---}
To overcome the main limitation of previous models, i.e., the full specification of the adjacency interaction matrix, we follow the approach used in statistical physics of spin systems called  \emph{simulated annealing} or \emph{configuration model} \cite{bianconi2002}.
We approximate $A_{ij}  \approx \frac{q_i^{(k)} q_j^{(k)}}{2 C^{(k)}}$, if $i$ and $j\in \mathcal{G}^{(k)}$. Here $q_i^{(k)}$ is the intra-group degree of node $i$ and $C^{(k)}$ is the total number of intra-group links in a group of size $k$. Similarly, $A_{ij} \approx  \frac{q_i^{(k,l)} q_j^{(l,k)}}{2 C^{(k,l)}}$ for $i \in \mathcal{G}^{(k)}$ and $j \notin \mathcal{G}^{(k)}$. Here  $q_i^{(k,l)}$ is the inter-group degree of node $i$ to all other groups of size $l$ and $C^{(k,l)}$ is the total number of inter-links between groups of size $k$ and $l$.
The simulated annealing approach can be understood as a dynamical friendship network formation framework based on the individuals' opinions and their desired number of friendship links.

\emph{Mean-field approximation.---}
Assuming the validity of a mean field approach, the group Hamiltonian can be approximated by $H_{MF}^{(k)}(\mathbf{s}_{i_1},\dots,\mathbf{s}_{i_k}) = \sum_{i \in \mathcal{G}^{(k)}} \mathbf{s}_i \cdot \mathbf{H}_{i}^{(k)}$, where
\begin{equation}
\mathbf{H}_i^{(k)} = -\frac{\phi J}{2} q_i^{(k)} \mathbf{m}^{(k)}  + \frac{(1-\phi) J}{2}  \sum_{l} q_\iota^{(k,l)} \mathbf{m}^{(l)} - h^{(k)}.
\end{equation}
We define $H_i^{(k)} = \mathbf{H}_i^{(k)} \cdot \mathbf{w}$.
By calculating the mean-field partition function and taking the derivative w.r.t. the external field, we get that the average group opinion, $m^{(k)}$, can be expressed as
\begin{equation}
m^{(k)} = \sum_{i \in \mathcal{G}^{(k)}} \tanh(\beta \, H_i^{(k)}(m^{(l)})) \, .
\end{equation}
{ A detailed derivation is found in the Supplemental Material (SM).} Let us now consider that the intra-group and inter-group degree distributions, $q^{(k)}$ and $q^{(k,l)}$, are random variables with  distributions $p(q^{(k)})$ and $\mathfrak{p}(\mathfrak{q}^{(k)})$, respectively. Then we can formulate the set of self-consistency equations
\begin{equation}
m^{(k)} = k \sum_{q^{(k)},q^{(k,l)}} p(q^{(k)}) p(q^{(k,l)}) \tanh(\beta H^{(k)})\, ,
\end{equation}
where $H^{(k)}$ depends on $q^{(k)}$, $q^{(k,l)}$ and $m^{(l)}$. Thus, we obtain a system of $n$ coupled equations for $m^{(k)}$.

\emph{Self-consistency equation with no inter-group interactions.---}
The set of self-consistency equations simplifies dramatically for the case $\phi=1$, where no inter-group interactions exist. The equations decouple, and we obtain one self-consistency equation for every $m^{(k)}$.
%
%
\begin{figure}[t]
    \centering
    \includegraphics[width=\linewidth]{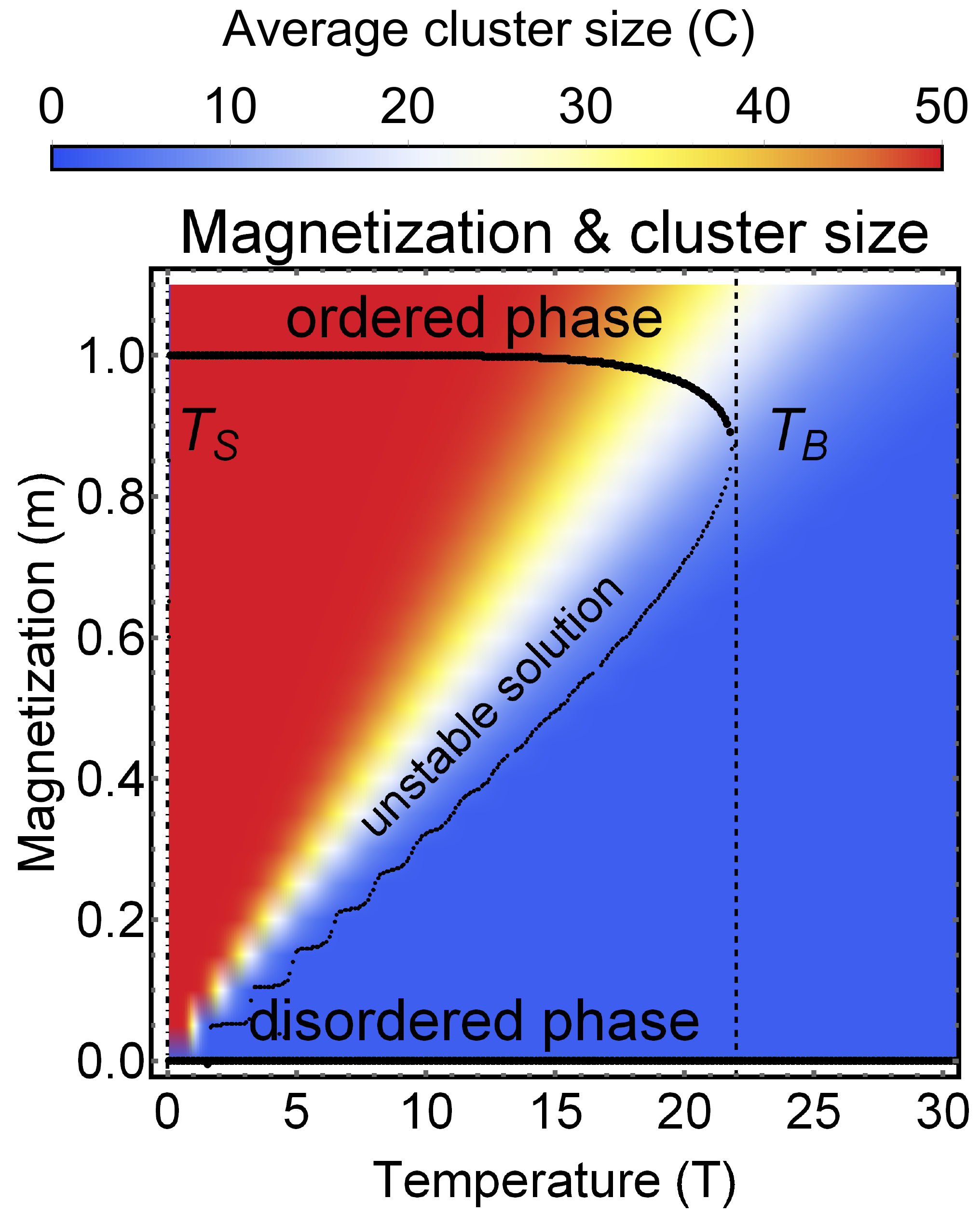}
    \caption{Total magnetization, $m$, as a function of the temperature, $T$, for $n=50$, $G=3$, and $J=1$, without external field. The heat map shows the average cluster size, $C$, as a function of $m$ and $T$. We observe the presence of a critical temperature, $T_B$, above which the self-consistency equations yield a single solution, $m=0$. Below this temperature, we observe two stable solutions, a coexistence of a disordered phase, characterized by the absence of large clusters (average cluster size is $1$), and an ordered phase that is characterized by the existence of one large cluster with all particles having the same opinion vector. By increasing the temperature to the critical temperature, the cluster starts to disintegrate rapidly while the magnetization remains relatively stable. }
    \label{fig:1}
\end{figure}
We first focus on the simple case of the fully connected inter-group network, where $p(q^{(k)}) = \delta(q^{(k)},k-1)$. In this case, the numerical value of the average magnetization per person is depicted in Fig. \ref{fig:1}. We observe the first-order phase transition between disordered 
phase
and the co-existence phase, where both disordered and ordered 
phase
exists. These phases are separated by the {\em binodal} temperature, $T_B$, which describes the point when the system, originally in the ordered phase, starts to disorder when increasing the temperature. Note that 
the {\em spinodal} temperature, $T_S$, tends to zero. The spinodal temperature describes the point where the particles spontaneously start forming large groups when starting in the disordered phase and decreasing the temperature. All individuals are free with random opinion in the disordered phase. In the ordered phase, all individuals form a single cluster with the same opinion. { The existence of the first-order transition between the ordered and a group phase has been suggested in  \cite{javarone2017evolutionary}.}

The average cluster size rapidly decreases near the binodal temperature while the overall magnetization remains relatively stable. However, at the critical temperature, the magnetization is still significantly non-zero
At the same time, the average cluster size decreases continuously toward one. The dependence of the phase diagram on the external field and minimum cluster size is shown in SM.

\begin{figure}
    \centering
    \includegraphics[width=\linewidth]{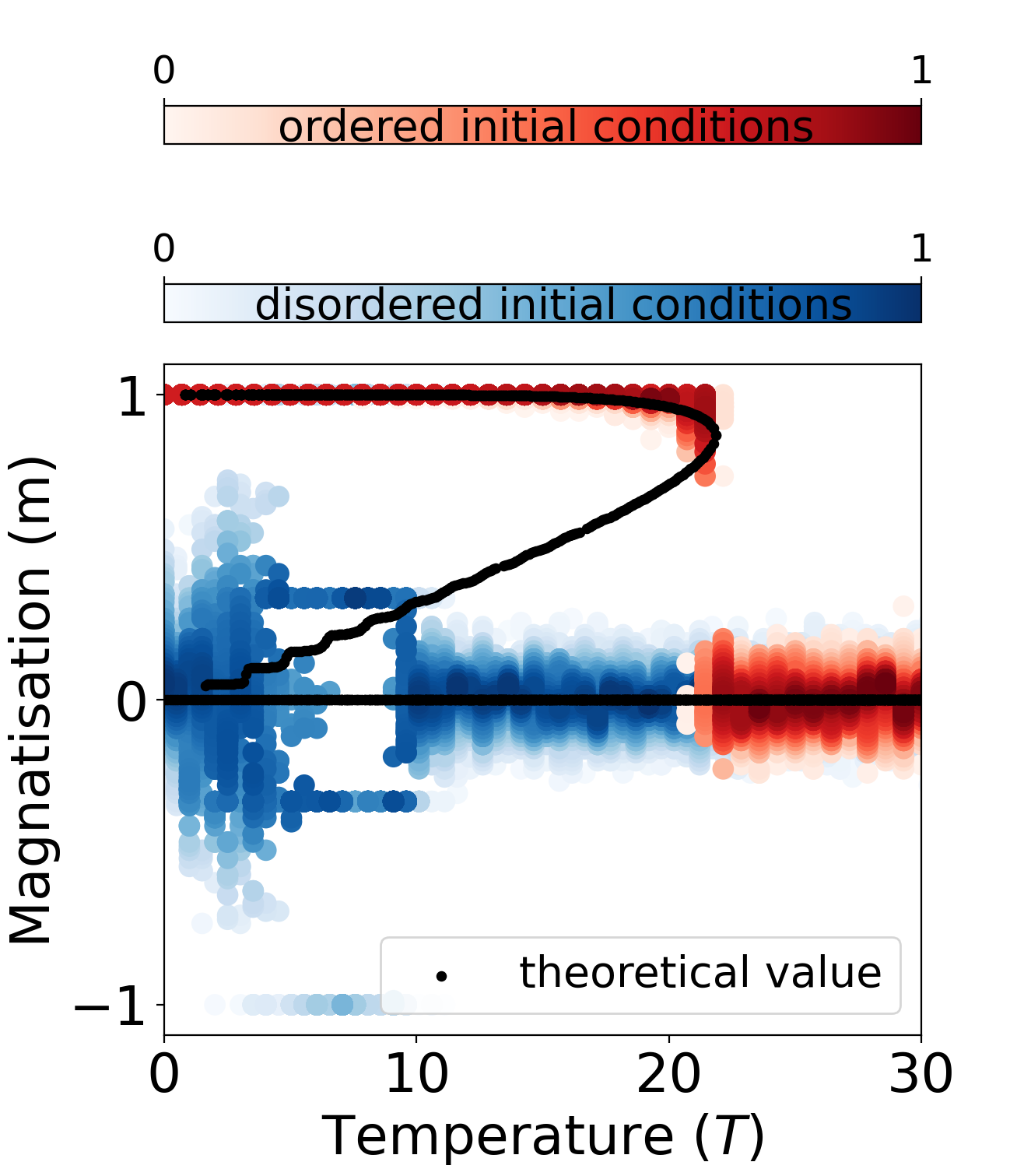}
    \caption{Magnetization as obtained from Monte Carlo simulations for the same parameters as in Fig. \ref{fig:1}. For each temperature, 100 independent runs with $5\cdot 10^4$ steps were performed. We started from two types of initial conditions --- one in the ordered phase, where all individuals are in one large group with identical initial opinions (red), and from the disordered phase, where all individuals form a separate group with a random initial opinion (blue). For each temperature, the histogram of magnetization was established. The darker the color, the larger the frequency; see color bars). The black curve shows the theoretical magnetization obtained from the self-consistency equations. For the ordered phase, we observe the perfect agreement with the theory, and the critical temperature corresponds with the predicted one. For the disordered phase, we see that the average temperature also corresponds to the predicted value, $m=0$. However, the fluctuations are larger. 
    }
    \label{fig:2}
\end{figure}

\emph{Monte Carlo simulations.---}
We perform the Monte Carlo simulations to confirm the phase diagram obtained from solving the self-consistency equations for the magnetization numerically. We use the standard Metropolis algorithm for $n=50$ individuals and three opinions, $G=3$. Each MC step consists of $n$ spin-updates (flips) of one spin-element chosen randomly, followed by randomly choosing one individual and moving them from the current group to another group -- or by creating a new group consisting of only that individual. If the individual is already solitary, it has to attach to one of the existing groups. 
The MC temperature for opinion flips and group changes are the same, i.e., both spin flips and group changes are accepted with probability, $\min\{1,\exp(-\beta \Delta H_{tot})\}$, where $H_{tot} = \sum_{\mathcal{G}^{(k)}} H(\mathcal{G}^{(k)})$ is the sum over all group Hamiltonians. For each temperature, we perform $100$ independent simulations with $5 \cdot 10^4$ MC steps. We repeat the simulation for two initial conditions corresponding to the two equilibrium phases: one in the ordered and one in the disordered phase.
For ordered initial condition, we observe that the particles stay in one cluster below critical temperature (Fig. \ref{fig:2} in red).
%
For the initial condition in the disordered phase,
the magnetization clearly fluctuates around zero (Fig. \ref{fig:2} in blue). For lower temperatures, the system can get stuck in a local minimum, seen by the fact that the magnetization fluctuates more and we observe a ``quantization'' effect.
Even for very low temperatures, large groups are rare to form and, therefore, $T_S$ is close to zero. In SM, we investigate the dependence of the phase diagram on the minimum group size, external field and initial conditions.

\begin{figure*}[t]
    \centering
    \includegraphics[width=\linewidth]{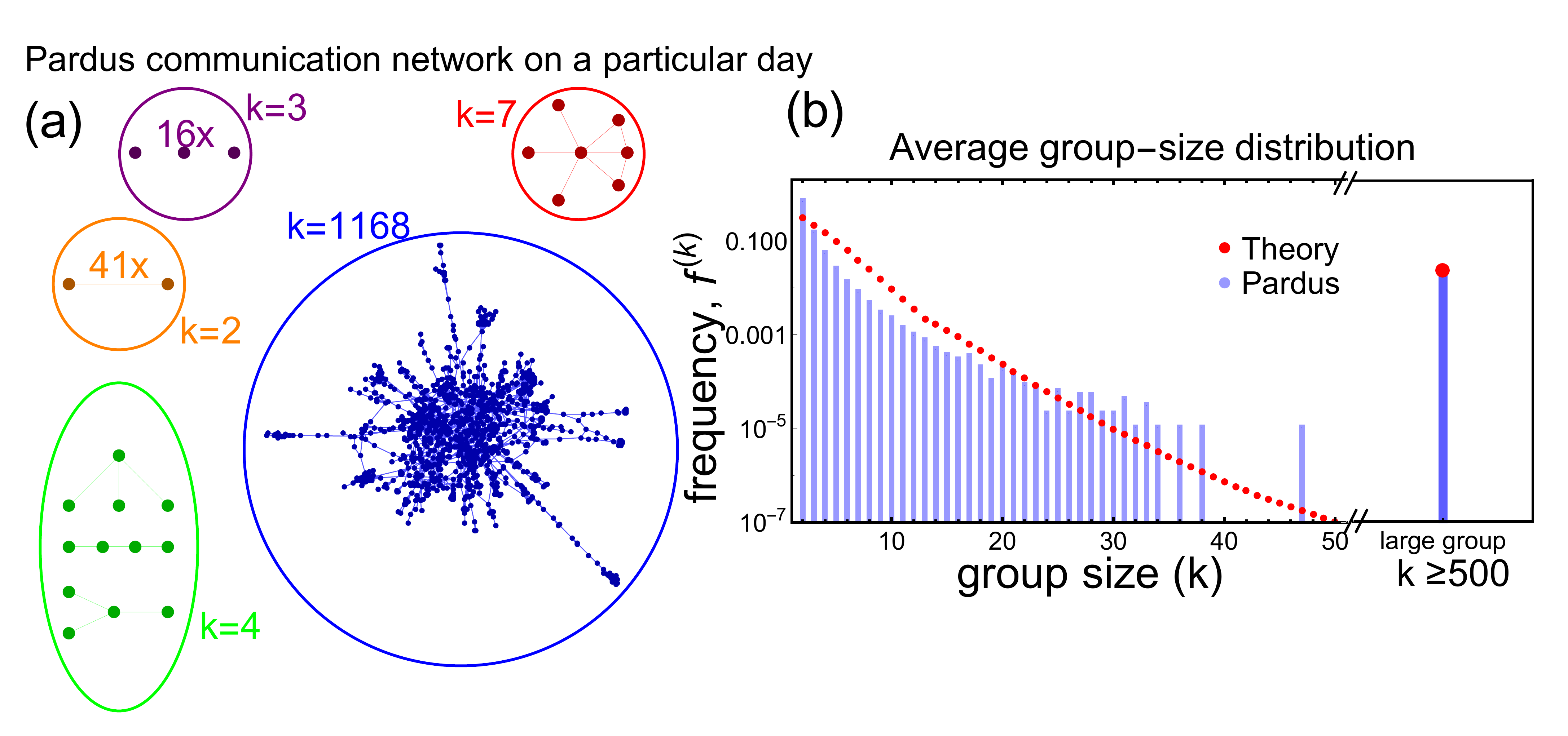}
    \caption{
    (a) an example of the  {\em Pardus}  communication network for the particular day 100. There is one giant connected component with a group size of $k=1168$ and several small components: one group of seven players, three groups of four players, 16 groups of three players, and 41 groups of two players.
    (b) Semi-logarithmic frequency distribution of group-sizes obtained from the  {\em Pardus}  dataset (blue, from \cite{Klimek2016}) and the prediction of the self-assembly group-formation model (red). The group size distribution is shown for group sizes between 2 and 50. For large groups of more than 500 players, the probabilities of observing a group with an exact particular size are very small. Thus we aggregate the probabilities for observing a group larger than 500 into one single bin. For both small and large groups, the theoretical prediction fits the group-size distribution of  {\em Pardus}  dataset well.}
    \label{fig:3}
\end{figure*}

\emph{Group-size distribution of Pardus network.---}
Finally, we compute the emerging group size distribution from the presented approach and compare it with a real dataset of an open-ended massive multiplayer online game called {\em Pardus} \cite{szell10}. Players in  {\em Pardus}  form friendships and enmity relations based on economic (in the virtual world) and social activities. We focus on only one type of social interaction, the players' communication. The dataset consists of 1239 days, each day with about 1000-1600 active players (players with at least one communication event with another player).
We adopt a picture where a communication event between players creates a link between them. Each connected component in this communication network corresponds to one group. Typically, we observe one giant connected component with several hundreds of participants and many small groups with sizes ranging from  $2$ to $50$. A typical communication network on one day is shown in Fig. \ref{fig:3} (a). The average group-size distribution is obtained by averaging group size distributions over all days.
We compare the so-obtained group-size distribution of the friendship network with the theoretical prediction from the self-assembly model in Fig. \ref{fig:3} (b). The intra-group degree distribution, obtained from \cite{Klimek2016}, can be well approximated with a truncated geometric distribution, $p^{(k)}_a(q^{(k)}) = a (1-a)^{q^{(k)}-1}/(1-(1-a)^k)$, where $q^{(k)} \in \{1,\dots,k-1\}$. 
with $a=0.6$, as shown in SM. For clarity, we show the frequency of observing a group of size $k$,  $f^{(k)} = n^{(k)}/M = n/M \cdot \wp^{(k)}$. By fitting the temperature, we obtain the theoretical group-size distribution, which corresponds to the real group-size distribution of the {\em Pardus}  dataset.
Due to the varying number of players across days, we fit the group-size distribution for small groups in the range between 2 and 50 and  aggregate the probability of observing one large group of more than 500 (giant component). Medium-size groups (51-499 players) do not appear in the dataset. It is obvious that the theoretical group-size distribution explains the empirical {\em Pardus} data well. { Interestingly, the Gini coefficient, which quantifies statistical dispersion, for both the empirical distribution ($G=0.900$) and the theoretical model ($G=0.901$) is close to the transition point of $G=0.86$ observed in several studies on percolation cluster size distribution (see \cite{Ghosh2022} for a recent review).}

\emph{Discussion.---}
The presented self-assembly model for social group formation offers a new view on
co-evolving
dynamics of group- and opinion formation. The framework of spin-glass self-assembly that is purely based on local information, i.e., local social stress and the number of contacts (degrees) of individuals.
Our main result is to show the existence of a critical temperature (binodal temperature) above which large groups disintegrate and that the opposite process, i.e., spontaneous group formation by lowering the temperature, is not possible when the external field is zero. 
We confirm these theoretical predictions with Monte Carlo simulations.

We are able to make a further testable prediction concerning the emerging group sizes in the model society. To compare with real data, using the social network of the {\em Pardus}  computer game, for which we have exact knowledge of group formation and sizes. Using the actual degree distribution of the friendship networks as inferred from the dataset as an input to our model, we are able to compute a group-size distribution that corresponds almost perfectly with the empirical group-size distribution in the  {\em Pardus} data.
{ Compared with recent work \cite{javarone2017evolutionary}, our study was able to take into account aspects such as social network topology, local homophily effects (not only the global average opinion), and co-evolution of opinions and the friendship links. Therefore, as a result, we obtained a complex phase diagram, including the previously observed first-order transition between individual and group phases and other phenomena, such as bifurcation of the average cluster size, or dependence on the external field. }

The model has a few limitations. The most important is that higher-order motives, such as those known from social balance, are not recovered correctly.
In many social networks, some of the higher-order motifs are  over- or underrepresented, compared to configuration model approach.
To get these statistics right, more advanced approaches known from spin glasses, such as the Bethe approach \cite{bethe1935statistical}, belief propagation \cite{yoon2011belief}, the cavity method \cite{mezard2003cavity}, or other generalizations of the configuration model could be useful.
A second limitation is a correlation between different opinions and the fact that this correlation can change over time. In reality, people define their \emph{belief system} where the opinions are correlated \cite{Rodriguez16,friedkin2016}, and the correlations can evolve in time. Finally, the presented framework operates on a single network where links between individuals influence all opinions of their friends/enemies. For more realism, one would need to consider a multilayer network that represents different environments (family, work, leisure time, social networks, etc.) with links of different types where each layer can influence only certain types of opinions. 
\begin{acknowledgments}
We thank Michael Szell and Peter Klimek for providing and assisting us with the Pardus data. The project was supported by the Austrian Science Fund (FWF) project No. P 34994 and project No. P 33751 and Austrian Science Promotion Agency (FFG) Project Grant 857136.
\end{acknowledgments}


\bibliography{bibliography}

\clearpage
\onecolumngrid
\appendix

\section*{Supplemental material}

{\subsection*{Detailed derivation of Eq. (4) in the main text}
In order to derive the self-consistency equation (4), we use the relation from the main text
\begin{equation}
m^{(k)} = - \frac{1}{\beta} \frac{\partial \log  \mathcal{Z}^{(k)}}{\partial h^{(k)}}
\end{equation}
To be able to calculate the partial partition function $\mathcal{Z}^{(k)}$, let's take the mean-field Hamiltonian
$H_{MF}(\mathbf{s}_{i_1},\dots,\mathbf{s}_{i_k}) = \sum_{i \in \mathcal{G}^{(k)}} \mathbf{s}_i \cdot \mathbf{H}_{i}^{(k)}$, where
\begin{equation}
\mathbf{H}_i^{(k)} = -\frac{\phi J}{2} q_i^{(k)} \mathbf{m}^{(k)}  + \frac{(1-\phi) J}{2}  \sum_{l} q_\iota^{(k,l)} \mathbf{m}^{(l)} - h^{(k)}.
\end{equation}
Thus, the partial partition function can be calculated as
\begin{eqnarray}
\mathcal{Z}^{(k)} &=& \frac{n^{k-1}}{k!} \sum_{\mathbf{s}_{i_1},\dots,\mathbf{s}_{i_k}} e^{-\beta H^{(k)}_{MF}(\mathbf{s}_{i_1},\dots,\mathbf{s}_{i_k})} = \frac{n^{k-1}}{k!} \sum_{\mathbf{s}_{i_1},\dots,\mathbf{s}_{i_k}} e^{-\beta \sum_{j \in \mathcal{G}^{(k)}} \mathbf{s}_j \cdot \mathbf{H}_{j}^{(k)}} = \frac{n^{k-1}}{k!} \prod_{j=1}^k \sum_{\mathbf{s}_j} e^{-\beta \mathbf{s}_j \mathbf{H}_j^{(k)}}\nonumber\\
&=&   \frac{n^{k-1}}{k!} \prod_{j=1}^k \prod_{l=1}^g \sum_{\mathbf{s}_j^l = \pm 1} e^{-\beta s_j^l H_j^l} =  \frac{n^{k-1}}{k!} \prod_{j=1}^k \prod_{l=1}^g \left(2 \cosh (\beta H_j^l\right))
\end{eqnarray}
Due to symmetry in $s_j^l$, we replace $H_j^l$ by its average over all opinions, i.e., $H_j^{(k)} = \mathbf{H}_j^{(k)} \cdot w$. Thus, the logarithm of the partition function can be expressed as
\begin{equation}
\log \mathcal{Z}^{(k)} = \ln \frac{2^{kG} n^{k-1}}{k!} +  \sum_{j=1}^k \ln  \cosh(\beta H_j^{(k)})
\end{equation}.
Note that $H_j^{(k)}$ depends on $m^{(l)} = \textbf{m}^{(l)}\cdot w$ and $h^{(k)}$. By taking the partial derivative w.r.t. $h^{(k)}$, and using the fact that $[\log(\cosh(x))]'= \tanh(x)$, we end with
\begin{equation}
m^{(k)} \equiv \frac{\partial \log \mathcal{Z}^{(k)}}{\partial h^{(k)}}   = \sum_{j \in \mathcal{G}^{(k)}} \tanh(\beta H_j^{(k)}(m^{(l)}))
\end{equation}
which corresponds to Eq. (4) in the main text.
}

\subsection*{Dependence of the phase diagram on the external field}
\begin{figure}[h]
    \includegraphics[width=0.49\linewidth]{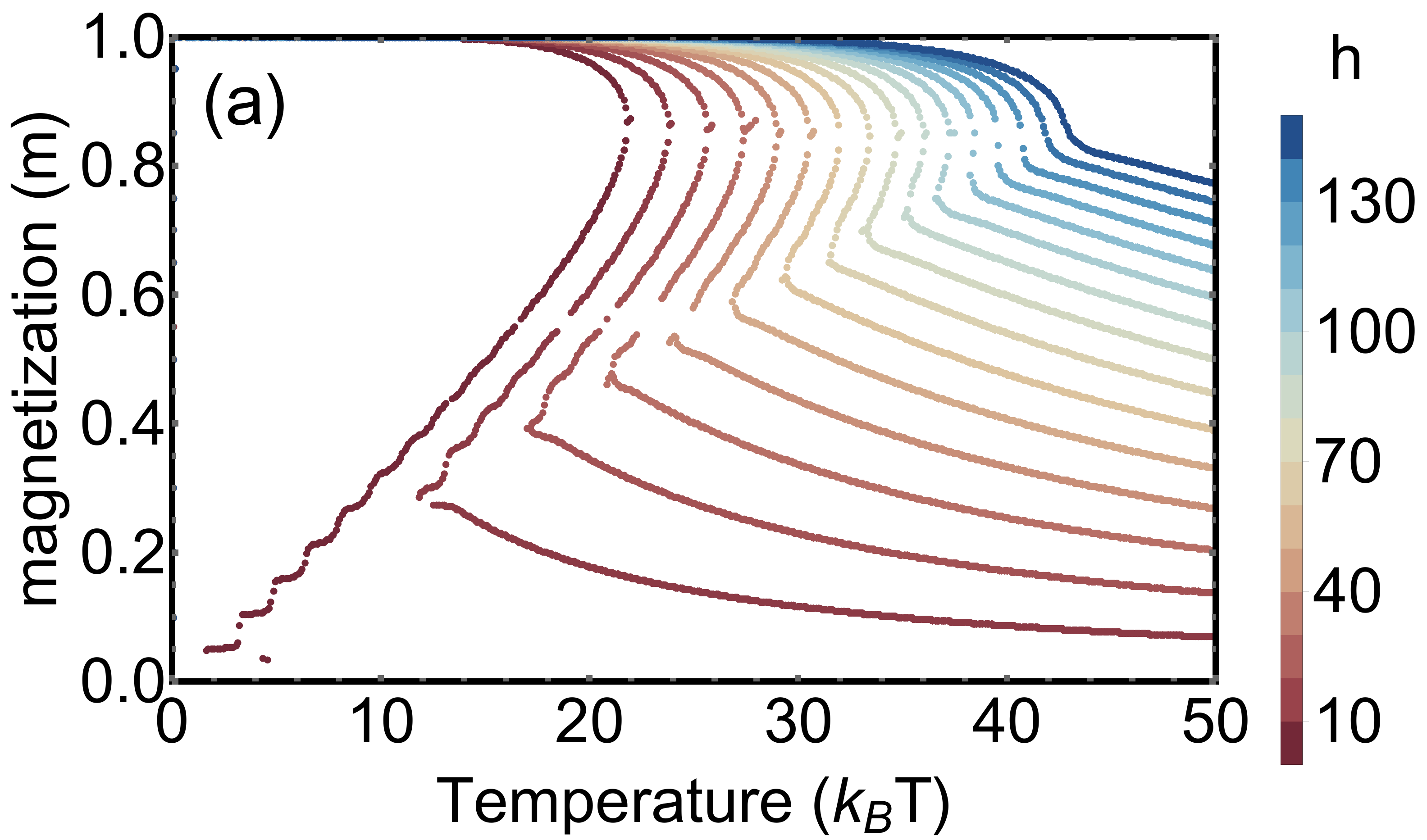}
    \includegraphics[width=0.40\linewidth]{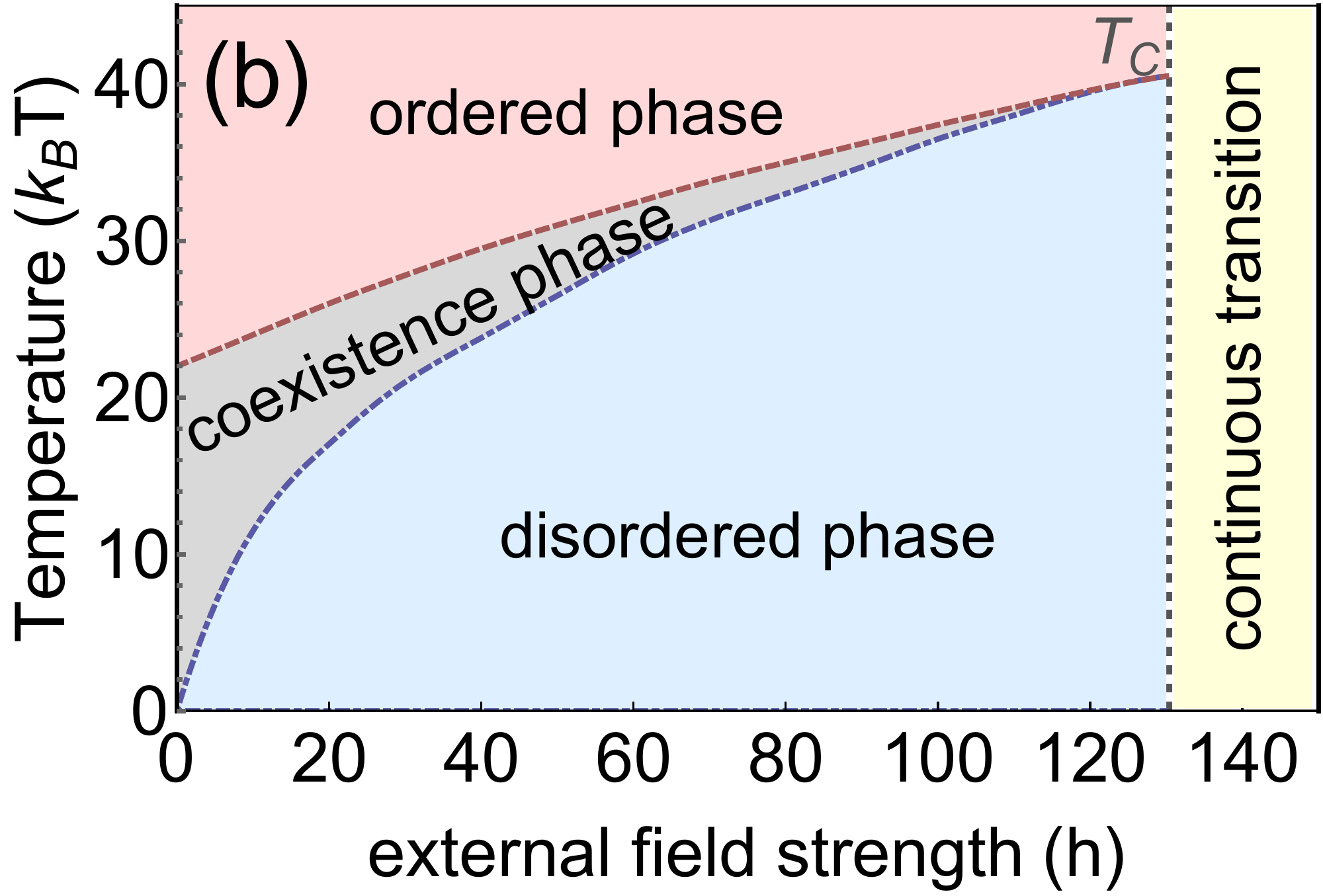}
    \caption{(a) Average magnetization $m$ for $n=50$, $G=3$, $J=1$ for various values of external field $h$. With increasing external field strength, the first-order transition in total magnetization $m$ changes to second-order and then to continuous transition between disordered and ordered phase. (b) The phase transition diagram is a function of temperature $T$ and external field $h$. Below the critical temperature $T_C$, we observe the first-order phase transition between the disordered and ordered phase and the existence of the coexistence phase. At the critical temperature, the transition becomes second-order. Above the critical temperature, we observe the super-critical phase, i.e., the continuous transition between two phases. }
    \label{fig:sm1}
\end{figure}

Let us now consider the case when the external field $h^{(k)} \equiv h$ is non-zero. Again, we numerically solve the self-consistency equation for the non-zero field. In Fig. \ref{fig:sm1} we observe that with increasing $h$, the phase transition changes from the first-order transition to the second-order transition at $h_C$ and then a becomes smooth transition. For $h < h_C$, we observe the bi-stable phase, which disappears at $T_C$. Thus, the external field plays the role of the transition parameter between the first-order and second-order transition and a smooth crossover, as described in \cite{kuehn2021universal}. Let us also note that we adhere to calling the phases ordered and disordered, although the disordered phase still exhibits partial order due to the presence of the magnetic field. However, the magnetization remains still smaller than in the case of the ordered phase.
One can interpret the presence of the external field as the influence of the mass media on individuals' opinions. By increasing the external field, one can force the system to switch to the ordered phase.

\subsection*{Dependence of the phase diagram on the minimal group size.}
\begin{figure}[h]
\centering
\includegraphics[width=0.49\linewidth]{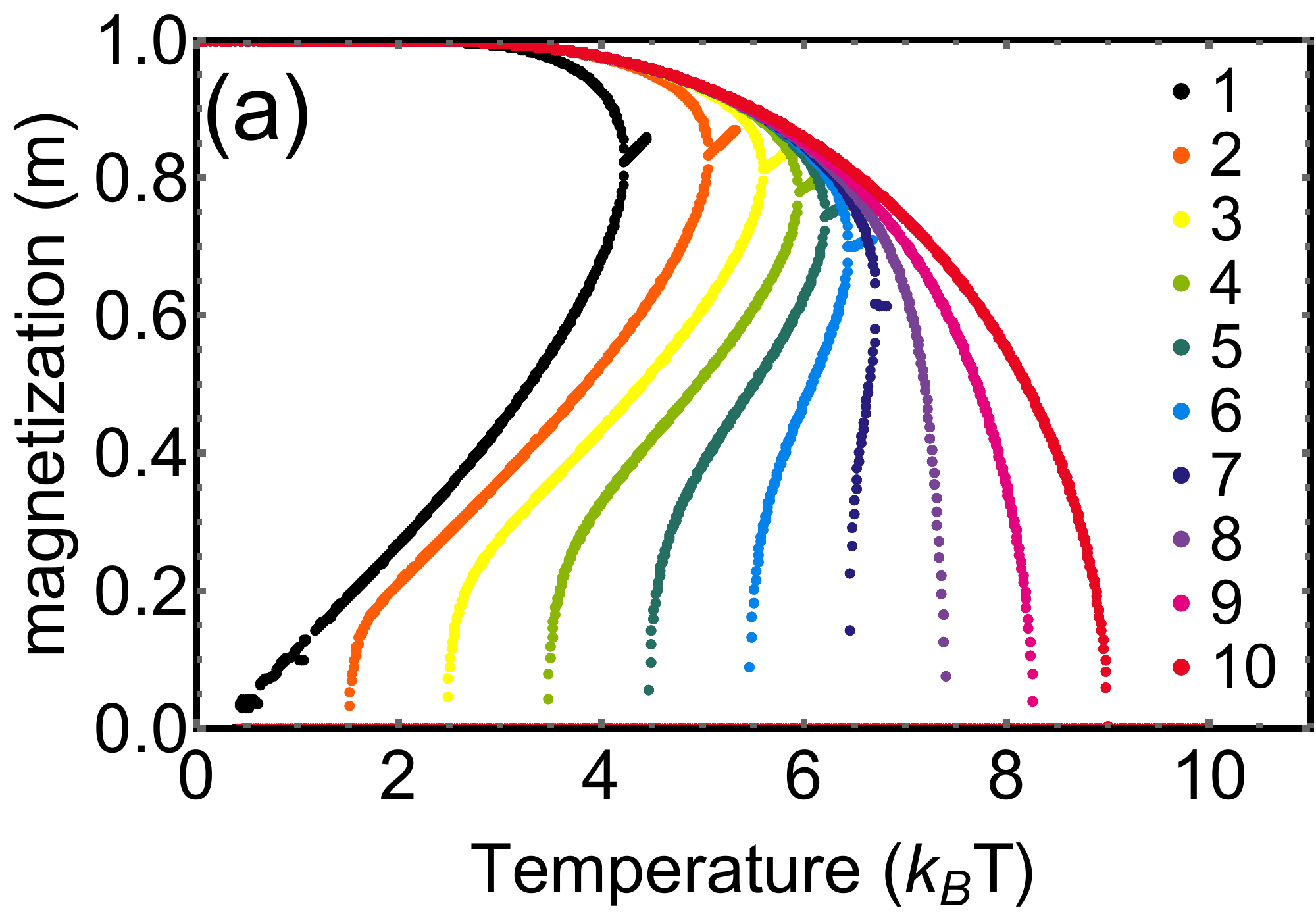}
\includegraphics[width=0.49\linewidth]{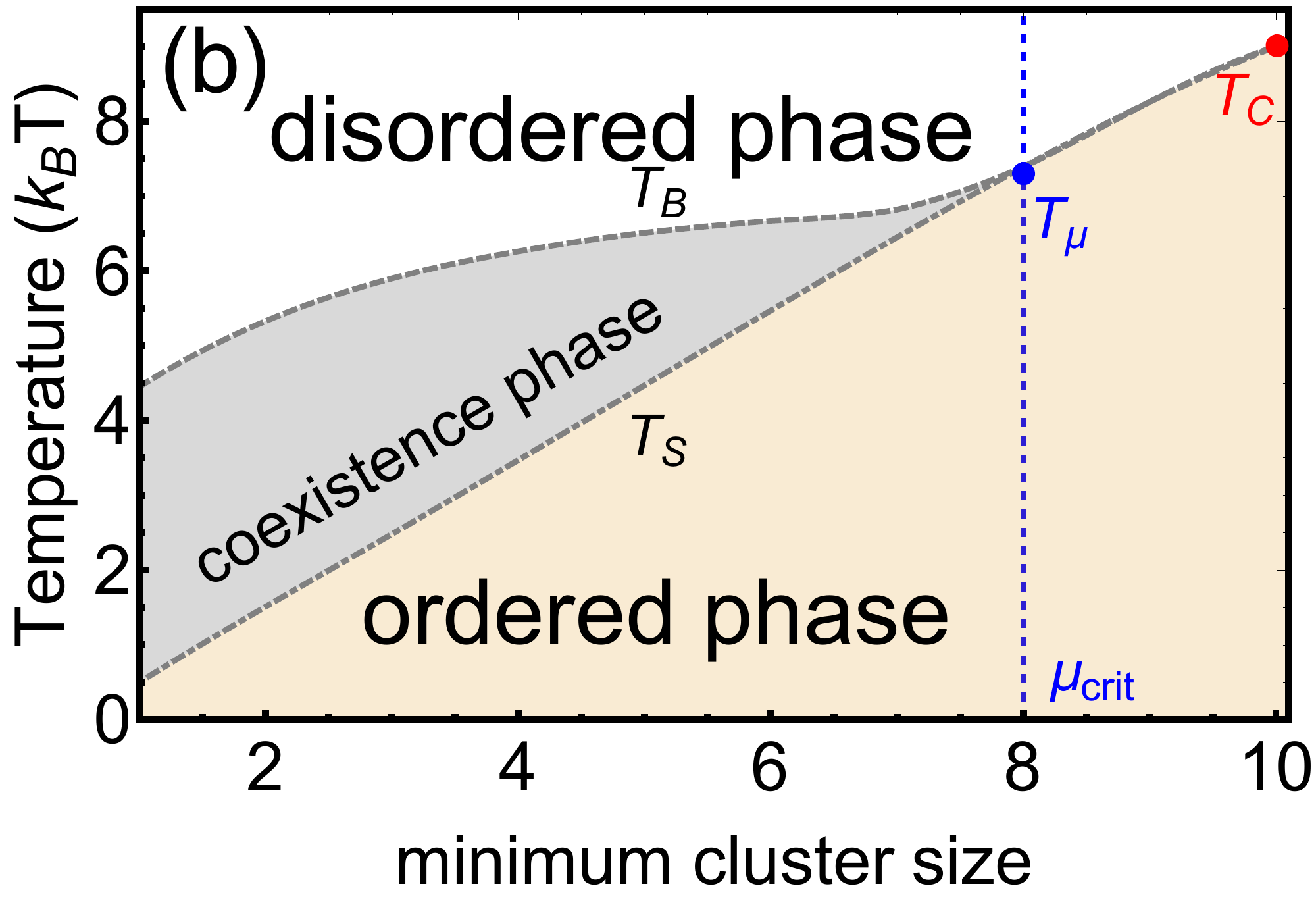}
\caption{(a) solution of the self-consistency equation for different minimum cluster sizes. The curves belonging to the minimum cluster size are depicted in different colors, corresponding to the legend. (b) Phase diagram of the temperature and minimum cluster size. We observe the existence of disordered phase and ordered phase. The coexistence phase exists when the minimum cluster size is smaller than the minimum size $m$. Above this threshold, the coexistence phase does not exist, and we observe a direct transition between ordered and disordered phases. For the case, when the minimum cluster size is equal to the maximum cluster size, both spinodal and bimodal temperature coincide with the Currie temperature of the Currie-Weiss model.}
\label{fig:sm2}
\end{figure}

In this case, we consider the case when the particles are not allowed to disintegrate fully, but they have to form a cluster of some minimum size. For each minimum size, we numerically calculate the solution of the self-consistency equation as shown on (a) of Fig. \ref{fig:sm2}. In this case, both spinodal and binodal temperatures grow with the increasing minimum size of the cluster. While the spinodal temperature grows linearly with the minimum cluster size, the binodal temperature growth slows down to the point where both temperatures coincide, and the co-existence phase disappears, see (b) of Fig. \ref{fig:sm2}. They both further grow together until the case where the minimum group size is equal to the number of particles, i.e., all particles belong to the same group. This corresponds to the original \emph{Currie-Weiss model}, and the critical temperature corresponds to Currie temperature $T_C = J (n-1)$.

\subsection*{Dependence of the Monte Carlo simulations on the initial conditions}
Let us focus on how the system relaxes with different initialization. All results are in Fig. \ref{fig:sm3}  First, we initialized the system randomly, i.e., each particle is assigned to a random group with a random spin. This corresponds to an ensemble given by the maximum entropy distribution with no constraint on average energy.
Moreover, we observe that in the case of random spin initialization, the particles can get stuck in a local minimum for a low temperature, not having enough energy to move to the next minimum. This is caused by the rough energy landscape (similarly to \cite{marvel2009}) caused by large energy differences when changing a group. Thus, the system can relax to the global minimum for a high enough temperature. The system disintegrates when the temperature further increases and gets into the disordered phase. In the third case, we initialize the system into two groups, and all spin equals $1$. Here we observe the existence of the \emph{hysteresis} region, where the system can relax to both ordered and disordered phases, depending on the exact trajectory, similarly to \cite{zhu2004hysteresis}. In the last case, we initialize the system into two groups of the same size and random spins. In this case, we observe all the aforementioned phenomena. With increasing temperature, we observe regions where the system gets stuck in a local minimum, ordered phase, hysteresis, and disordered phase.
\begin{figure*}[h]
\includegraphics[width=0.32\linewidth]{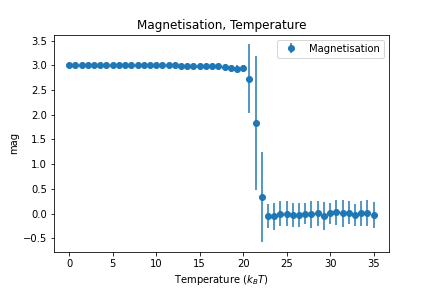}
\includegraphics[width=0.32\linewidth]{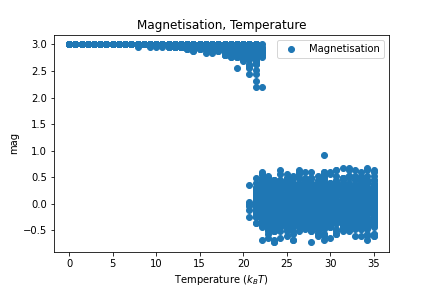}
\includegraphics[width=0.32\linewidth]{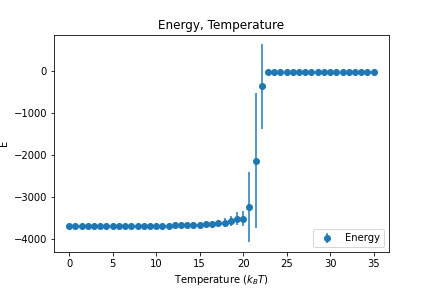}\\

\includegraphics[width=0.32\linewidth]{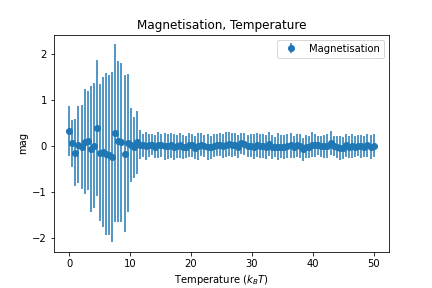}
\includegraphics[width=0.32\linewidth]{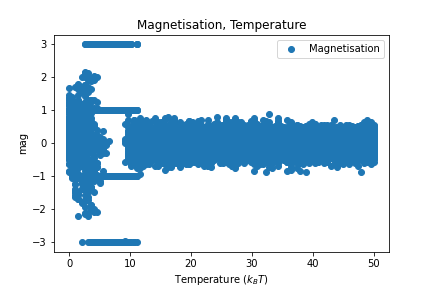}
\includegraphics[width=0.32 \linewidth]{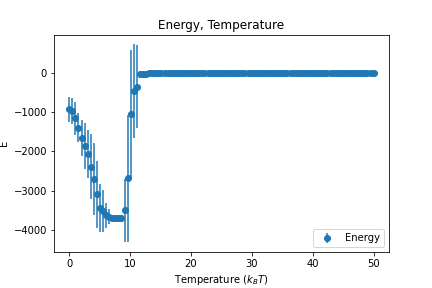}\\
\includegraphics[width=0.32\linewidth]{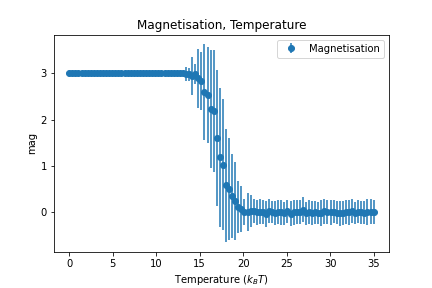}
\includegraphics[width=0.32\linewidth]{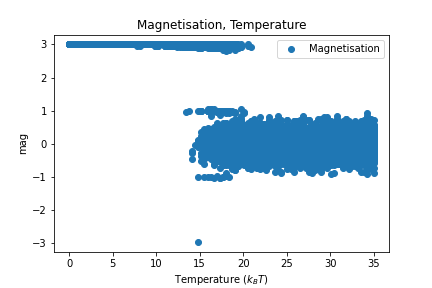}
\includegraphics[width=0.32\linewidth]{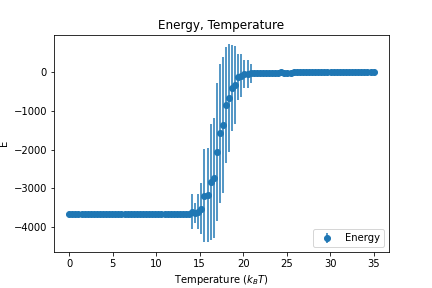}
\includegraphics[width=0.32\linewidth]{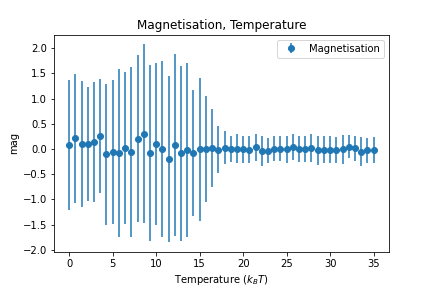}
\includegraphics[width=0.32\linewidth]{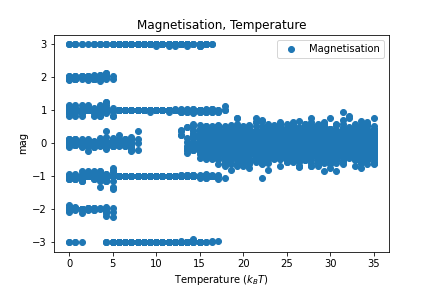}
\includegraphics[width=0.32\linewidth]{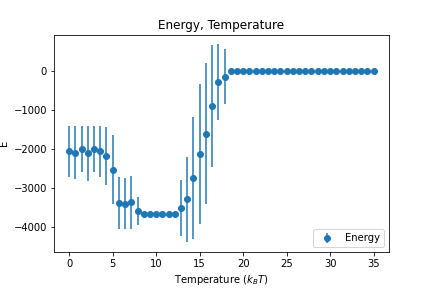}
\caption{Monte Carlo Simulations of self-assembly of Ising particles. The simulations are done for $n=50$, $G=3$ and $J=1$. Make $5 \cdot 10^4$ MC steps for each temperature and perform $100$ runs. We plot the error plot of final magnetization (left), the scatter plot of final magnetization (center), and the error plot of final energy (right). In the first row, we started with all particles in one cluster with all spins $1$. In the second row, we started with random initialization of the system. In the third row, we started with two groups of the same size and all spins $1$, and finally, in the last row, we started with the two groups of the same size with random initial spins.}
\label{fig:sm3}
\end{figure*}

\subsection*{Degree distribution of Pardus and fit by truncated geometric distribution}
As mentioned in the main text, can be well approximated with a truncated geometric distribution, $p^{(k)}_a(q^{(k)}) = a (1-a)^{q^{(k)}-1}/(1-(1-a)^k)$, where $q^{(k)} \in \{1,\dots,k-1\}$. For Pardus, we obtain $a=0.6$ gives the decent fit for all group sizes. This is illustrated in Fig. \ref{fig:sm3} for the group size $n=10,20,30$, but we obtain similar fits for all other group sizes.
\begin{figure*}
\includegraphics[width=\linewidth]{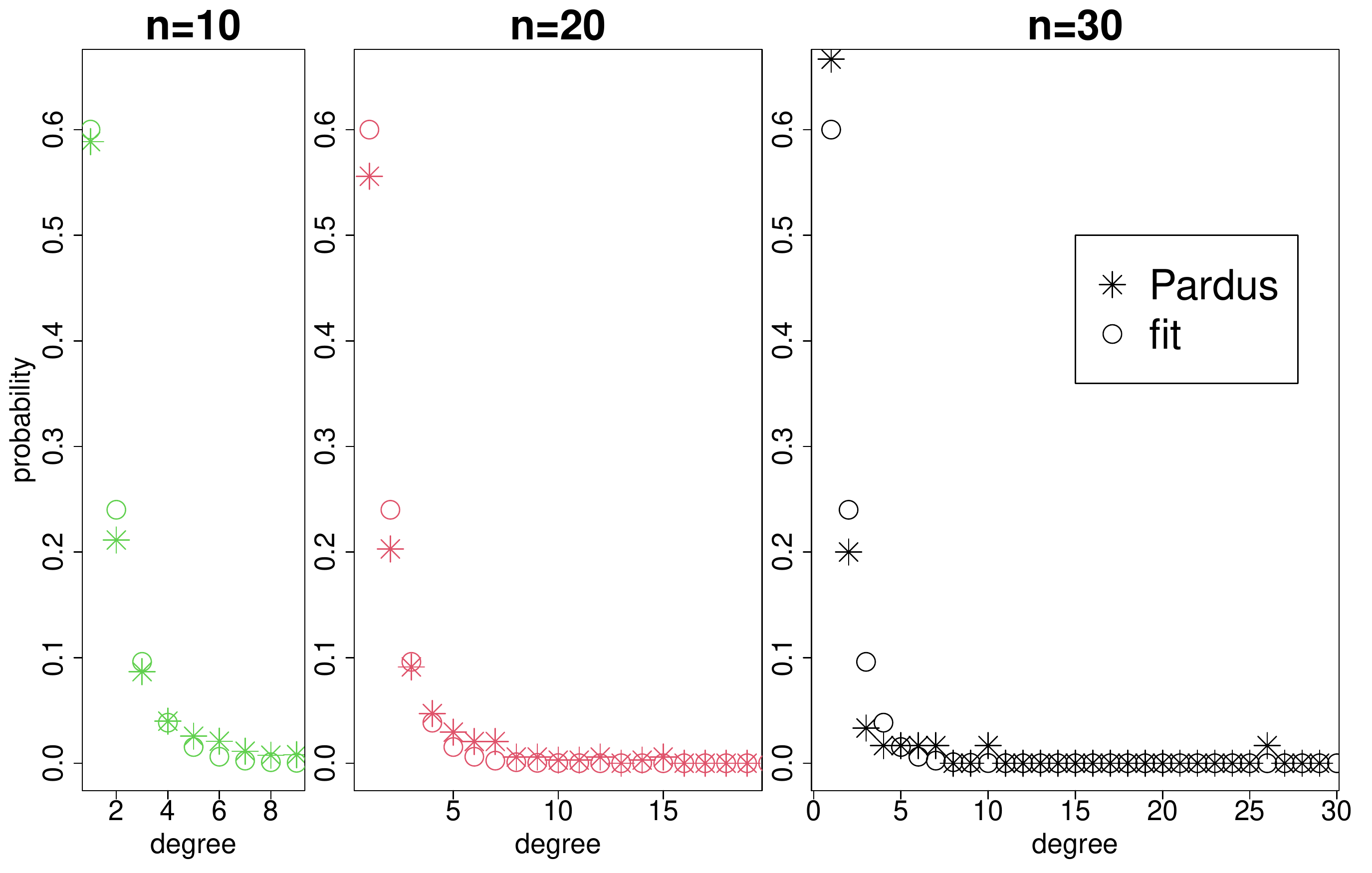}
\caption{Degree distribution of the pardus dataset for group sizes $n=10,20,30$ and the corresponding truncated geometric distribution for $a=0.6$.}
\label{fig:sm3}
\end{figure*}
\bibliography{bibliography}

\end{document}